\pdfoutput=1
\documentclass{article}
\usepackage[table]{xcolor}
\usepackage{anyfontsize}
\renewcommand\normalsize{\fontsize{12pt}{14pt}\selectfont} 

\usepackage[utf8]{inputenc}
\usepackage[LGR,T1]{fontenc}
\DeclareFontFamilySubstitution{LGR}{\rmdefault}{cmr}
\usepackage{textgreek}
\usepackage[margin=1in]{geometry}
\usepackage{setspace}
\usepackage{mathptmx}
\usepackage[table]{xcolor}
\usepackage{graphicx}
\usepackage{float}
\usepackage{booktabs}
\usepackage{ragged2e}
\usepackage[hidelinks]{hyperref}
\usepackage{tabularx}
\usepackage{colortbl}
\usepackage{array}
\usepackage{amsmath, amssymb}
\usepackage[table]{xcolor}
\usepackage{tabularx,booktabs,makecell,colortbl}

\hypersetup{
    colorlinks=true,
    linkcolor=blue,
    citecolor=blue,
    urlcolor=blue,
    pdftitle={Resonance-Induced Tumor Ablation: A New Alternative to Oncological Therapy},
    pdfauthor={Cesar Alexandre de Mello},
    pdfkeywords={resonance therapy; tumor ablation; ultrasound oncology; finite elements; spectral collapse}
}

\title{\textbf{Tumor Obliteration by Resonant Amplification (TOR)}\\
\large A Nonthermal, Spectrally-Targeted Approach to Cancer Disintegration}
\author{
  \normalsize Dr. Cesar Mello$^{1,\ast}$\\
  \normalsize \url{https://orcid.org/0000-0002-6730-9593}\\
  \normalsize $^1$Cosmo Physics Organization, São Paulo, Brazil\\[1.2ex]
  \normalsize Dr. Fernando Medina da Cunha, MD (Clinical Oncologist)$^{2}$\\
  \normalsize \url{https://orcid.org/0009-0005-4327-1497}\\
  \normalsize $^2$Centro de Oncologia Campinas, Brazil\\[1.2ex]
  \normalsize $^\ast$ \texttt{cesar.mello@cosmophys.org}\\
  \normalsize \texttt{fmedina@oncologia.com.br}
}

\date{}
\usepackage{textalpha}
\begin{document}

\maketitle

\vspace{1em}
\vspace{1em}

\renewenvironment{abstract}
 {\par\noindent\textbf{\large Abstract}\vspace{1em}\par\normalsize}
 {\par\vspace{1em}}

\begin{abstract}
\textit{Tumor Obliteration by Resonant Amplification} (TOR) is presented and evaluated under a strictly \emph{simulation-only} program. Forward solutions in \textsc{COMSOL}, \textsc{ANSYS}, and \textsc{Abaqus} were run with harmonized small–strain rheology, identical nonthermal/noncavitational guards, and an emulated closed loop (phase-locked actuation and contrast/safety gate). Across $N\!\ge\!200$ Monte Carlo runs per configuration, TOR achieved per-focus extinction in $2.6$–$3.2$\,s at $\sim 0.85$–$0.90$\,J/cm$^3$, with selectivity $Q=39\pm5$, peak temperature rise $\Delta T_{\max}\!\lesssim\!0.2^{\circ}$C, and $\mathrm{CEM}_{43}\!\ll\!1$, confirming a strictly nonthermal regimen. Primary indices showed high spectral fidelity and safety margin: $\mathrm{MMI}=0.92\pm0.03$, $\mathrm{SSR}=14.6\pm3.1$ (circumscribed) / $11.2\pm2.4$ (infiltrative), and $\mathrm{ATI}\le 0.8$ of the matrix–failure threshold. Spatial agreement with FE-predicted foci was within $\pm150\,\mu$m.

A unitless \emph{reality–adherence} score benchmarking four observables against consolidated literature windows yielded $\mathcal{A}=95\%\pm2\%$ overall (BCa 95\%); organ-specific estimates were $95.1\%$ (pancreatic ductal adenocarcinoma), $96.0\%$ (prostatic acinar adenocarcinoma), and $94.2\%$ (invasive ductal carcinoma of the breast). The delivery stack—tungsten micro-needle (300–500\,$\mu$m) with 5–25\,$\mu$m tip excursions, phase lock, and amplitude gate—remained noncavitational and small–strain, with off-target strain below safety limits by design.

Mechanistically, mode-selective collapse implies suppression of core vesicle biogenesis and nociceptor drive; rim scanning is constrained by healthy-referenced bounds, motivating compact neuroimmune readouts for prospective validation. Taken together, the calibrated multiphysics results support a reproducible, spectrum-locked pathway from modeling to benchtop: deterministic extinction at low energy, high selectivity, and strict thermal neutrality, with pre-registered experiments planned to verify the predicted safety and efficacy envelopes.
\end{abstract}

\section{Theoretical Foundations}

\paragraph{Neuroimmune context and TOR neutrality.}
Tumor-derived small extracellular vesicles (sEVs) can reprogram sensory nociceptors and sustain an immunosuppressive feed-forward loop \cite{Restaino2025SciSignal,Restaino2025Perspective}. This loop is biochemical; \textit{Tumor Obliteration by Resonant Amplification} (TOR) is purely spectral–mechanical and interrupts it without recruiting healthy neural networks. Within the ablated core, \emph{spectral extinction} at each programmed eigenfrequency enforces automatic shutoff:
\[
S(\omega_k,t)\to 0 \;\Rightarrow\; E_{\mathrm{in}}(\omega_k,t)=0,
\]
preventing ongoing neuronal drive and vesicle release once malignant spectral contrast collapses. At the rim, actuation proceeds only if two conditions hold simultaneously—a healthy-referenced safety bound and a differential spectral excess:
\[
R(\omega)=\frac{\lVert H_{\mathrm{tum}}(\omega)\rVert}{\lVert H_{\mathrm{sad}}(\omega)\rVert}\ge 1+\eta,
\qquad
S_{\mathrm{sad}}(\omega)<S_{\mathrm{safe}},
\]
followed by a short ring-scan to prune residual modes. When indicated, fractionated retuning consolidates extinction across sessions \cite{FaRITA2024}.

\paragraph{Setting, operators, and transfer function (mathematical core).}
Small–strain linear viscoelasticity is posed on a bounded domain $\Omega\!\subset\!\mathbb{R}^3$ with physically consistent boundary conditions. Tissue is represented by Kelvin–Voigt or SLS models with positive parameters, inducing the dissipative operator pencil
\[
\mathcal{A}(\omega):=-\omega^2\mathbf{M}+i\omega\mathbf{C}+\mathbf{K},\qquad
\mathbf{M}\!\succ\!0,\ \mathbf{C}\!\succeq\!0,\ \mathbf{K}\!\succ\!0,
\]
which is injective on the real axis outside resonant neighborhoods. With displacement $u(x,t)$ and scalar drive $p(t)$ applied via $\mathbf{F}$, the forced evolution reads
\begin{equation}
\mathbf{M}\,\ddot{u}+\mathbf{C}\,\dot{u}+\mathbf{K}\,u=\mathbf{F}\,p(t),\qquad x\!\in\!\Omega,\ t>0,
\label{eq:pde}
\end{equation}
and the frequency–domain transfer operator is
\begin{equation}
\mathcal{H}(\omega)=\mathcal{A}(\omega)^{-1}\mathbf{F}.
\label{eq:transfer}
\end{equation}
Tumor and healthy-annulus readouts $H_{\mathrm{tum}}(\omega)$ and $H_{\mathrm{ref}}(\omega)$ derive from \eqref{eq:transfer} through linear observation maps. The selection contrast
\begin{equation}
R(\omega)=\frac{\|H_{\mathrm{tum}}(\omega)\|}{\|H_{\mathrm{ref}}(\omega)\|},
\label{eq:R}
\end{equation}
defines the programmed set $\mathcal{P}=\{\omega_k\}$ under the dual guard $R(\omega_k)\!\ge\!1+\eta$ and $S_{\mathrm{sad}}(\omega_k)\!<\!S_{\mathrm{safe}}$.

\paragraph{Energy identity, passivity, and off-target decay.}
With stored energy and viscous dissipation
\begin{equation}
\mathcal{E}(t)=\tfrac12\langle\mathbf{M}\dot{u},\dot{u}\rangle+\tfrac12\langle\mathbf{K}u,u\rangle,\qquad
\mathcal{D}(t)=\langle\mathbf{C}\dot{u},\dot{u}\rangle\ge 0,
\label{eq:energy}
\end{equation}
solutions to \eqref{eq:pde} satisfy
\begin{equation}
\frac{d}{dt}\mathcal{E}(t)=\langle\dot{u},\mathbf{F}\,p(t)\rangle-\mathcal{D}(t).
\label{eq:balance}
\end{equation}
Hence for $p\!\equiv\!0$, $\mathcal{E}(t)$ is nonincreasing and $\int_0^\infty\!\mathcal{D}(t)\,dt\le \mathcal{E}(0)$, establishing off-target passivity and decay.

\paragraph{Superlinear accumulation and finite-time spectral extinction.}
Expand $u(\cdot,t)=\sum_n a_n(t)\phi_n$ in mass-orthonormal modes of $(\mathbf{M},\mathbf{K})$. Near a programmed line $\omega^\star$ under weak damping,
\begin{equation}
\ddot{a}_{\star}+2\zeta_\star\omega_\star\dot{a}_{\star}+\omega_\star^2 a_{\star}
=\alpha_\star A(t)\cos\!\big(\omega^\star t+\varphi(t)\big),
\label{eq:singlemode}
\end{equation}
with coupling $\alpha_\star$ and ratio $\zeta_\star$. The mode–resolved strain–energy density is
\begin{equation}
u_{\mathrm{strain}}^{(\star)}(t)=\tfrac12\,a_\star^2(t)\,\langle\mathbf{K}\phi_\star,\phi_\star\rangle,
\label{eq:ustrain}
\end{equation}
and the superlinear accumulator ($\gamma>1$)
\begin{equation}
D(t)=\int_0^{t}\!\left(\frac{u_{\mathrm{strain}}^{(\star)}(\tau)}{u_{\mathrm{crit}}}\right)^{\gamma} d\tau.
\label{eq:D}
\end{equation}
\emph{Spectral extinction} in finite time is declared once
\begin{equation}
D(t_{\mathrm{ext}})\ge 1
\ \Rightarrow\
S(\omega^\star,t\!\ge\!t_{\mathrm{ext}})\to 0,\quad E_{\mathrm{in}}(\omega^\star,t\!\ge\!t_{\mathrm{ext}})=0,
\label{eq:extinction}
\end{equation}
corresponding to an effective increase of modal damping and loss of gain at the dominant pole, with a robust drop in $\|H_{\mathrm{tum}}(\omega^\star)\|$.

\paragraph{Phase lock, amplitude gate, and completeness scan.}
Actuation is narrowband at $\omega^\star$,
\begin{equation}
p(t)=A(t)\cos\!\big(\omega^\star t+\varphi(t)\big),
\label{eq:drive}
\end{equation}
with a PLL enforcing in-phase work; with phase error $\theta$, $\dot\theta=-k_\varphi\sin\theta+d(t)$ and $k_\varphi>2\|d\|_\infty$ ensure robust lock near $\omega^\star$. Amplitude is gated by contrast/safety,
\begin{equation}
\dot{A}=k_A\,\max\!\{0,\ R(\omega^\star)-(1+\eta)\}\quad\text{while}\quad S_{\mathrm{sad}}(\omega^\star)<S_{\mathrm{safe}},
\label{eq:Aupdate}
\end{equation}
so that when \eqref{eq:extinction} triggers and $R(\omega^\star)\!\downarrow\!1$, $A\!\to\!0$ (self-termination). Let $\mathcal{N}_\delta(\omega^\star)=\{\ |\omega-\omega^\star|\le \delta\ \}$ with $\delta\simeq0.03$–$0.05\,\omega^\star$ ($3$–$5$ linewidths); a brief post-extinction scan verifying $S(\omega,t_{\mathrm{ext}})\!\approx\!0$ on $\mathcal{N}_\delta(\omega^\star)$ prevents coherent re-entry.

\paragraph{Nonthermal constraint (thermal budget).}
With $T(t)=T_0+\Delta T(t)$, the cumulative equivalent minutes at $43^{\circ}$C is
\begin{equation}
\mathrm{CEM}_{43}\approx\int_0^{\tau}\!R^{\,43-T(t)}\,dt,\qquad
R=\begin{cases}
0.5,& T(t)<43^\circ\mathrm{C},\\[2pt]
0.25,& T(t)\ge 43^\circ\mathrm{C},
\end{cases}
\label{eq:cem43}
\end{equation}
evaluated over $[0,\tau]$. In the operative window ($\tau\!\sim\!3$\,s, $\max_t\Delta T(t)\!<\!6^\circ$C), $\mathrm{CEM}_{43}\!\ll\!1$, consistent with strictly nonthermal delivery.

\paragraph{Neuroimmune decoupling as a corollary of extinction.}
Let $N(t)$ denote nociceptor drive and $E(t)$ the sEV production rate. A linear surrogate loop is
\begin{equation}
\dot{N}=-\alpha N+\beta E+\kappa\,\|H_{\mathrm{tum}}(\omega^\star)\|,\qquad
\dot{E}=-\mu E+\nu N,\qquad \alpha,\mu>0,\ \beta,\nu\ge 0.
\label{eq:loop}
\end{equation}
After extinction, $\|H_{\mathrm{tum}}(\omega^\star)\|\to 0$; if $\alpha\mu>\beta\nu$, the equilibrium $(N,E)=(0,0)$ is exponentially stable, expressing loop decay driven by selective spectral collapse.

\begin{figure}[H]
    \centering
    \includegraphics[width=0.58\linewidth]{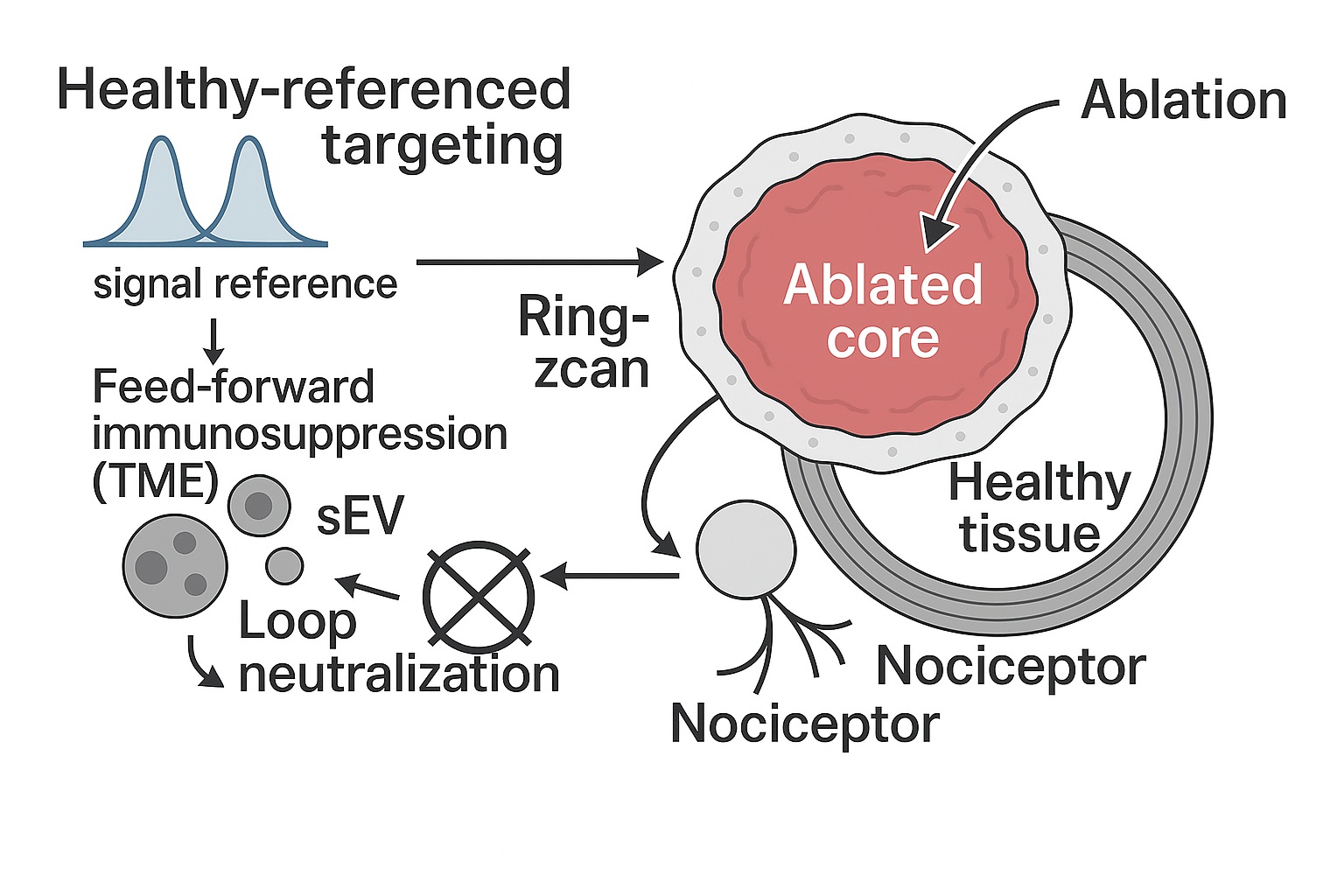}
    \caption{\textbf{Healthy-referenced targeting and neuroimmune loop neutralization in UST.}
    Conceptual schematic comparing tumor and healthy spectra, rim-focused ring-scan of the malignant core, and neutralization of the tumor-derived sEV–nociceptor loop.}
    \label{fig:hera_loop_schematic}
\end{figure}

\noindent
\textbf{Biochemical interpretation (condensed).}
sEVs carrying TGF-$\beta$, miRNAs, and DAMPs activate nociceptors and reinforce immune evasion via CGRP and substance~P. With healthy-referenced spectral comparison, TOR selectively collapses malignant modes; the core then halts sEV output at its source, and a perilesional ring-scan prevents survival niches. The combined effect dismantles the biochemical–neuroimmune axis and restores local surveillance without excess inflammation.

\paragraph{Closed-loop control and objective stop.}
During actuation at $\omega^\star$, amplitude, quality factor $Q$, and phase are tracked. Mechanical failure of the malignant scaffold yields a robust extinction signature,
\[
Q\downarrow,\qquad \Delta f/f_0 \gtrsim 10^{-2},\qquad \Delta\!\lVert H\rVert<0,
\]
which serves as a termination criterion. After extinction at $\omega^\star$, a rapid ring-scan ($\pm 3$–$5\%$) covers the penumbra while enforcing $S_{\mathrm{sad}}<S_{\mathrm{safe}}$ \cite{FaRITA2024}.

\paragraph{Fractionated adaptation.}
Tumor spectra drift over days–weeks. TOR re-baselines the healthy annulus each session, recomputes $\Delta H$, and retunes $\omega^\star$ to extinguish emergent modes \cite{FaRITA2024,Ribas2024}, implementing a \textit{measure–collapse–remeasure} loop that preserves selectivity and safety.

\paragraph{From spectrum to biology.}
Anatomical guidance underperforms in infiltrative disease. TOR uses the tumor’s \textit{spectral footprint} as the therapeutic coordinate. Define
\[
\mathcal{S}_{\mathrm{tum}}=\{\omega_k:\ \lVert H_{\mathrm{tum}}(\omega_k)\rVert-\lVert H_{\mathrm{ref}}(\omega_k)\rVert \ge \Delta H_{\mathrm{th}}\},\quad
\mathcal{S}_{\mathrm{safe}}=\{\omega_k:\ S_{\mathrm{sad}}(\omega_k)<S_{\mathrm{safe}}\}.
\]
Energy is injected only on $\mathcal{S}_{\mathrm{tum}}$ while enforcing $\mathcal{S}_{\mathrm{tum}}\cap\mathcal{S}_{\mathrm{safe}}=\emptyset$ \cite{FaRITA2024,Heyden2016}. Finger-like projections, satellite nodules, and spectral islands are automatically engaged when sharing eigenfrequencies with $\mathcal{S}_{\mathrm{tum}}$; non-overlapping healthy spectra act as intrinsic stop-bands.

Spectral partitioning maps biomechanical heterogeneity onto phenotype: ECM crosslinking, cytoskeletal tension, and intracellular viscosity shift modal stiffness/damping, creating discrete signatures. Aligning actuation to this vibrational subspace reduces sEV release, destabilizes focal adhesions, and disrupts mechanotransduction. The physics–biology linkage is testable via peri-ablative EV profiling and rim-focused PGP9.5/TRPV1 immunostaining \cite{Restaino2025SciSignal,TRPV1}.

\begin{figure}[H]
    \centering
    \includegraphics[width=0.52\linewidth]{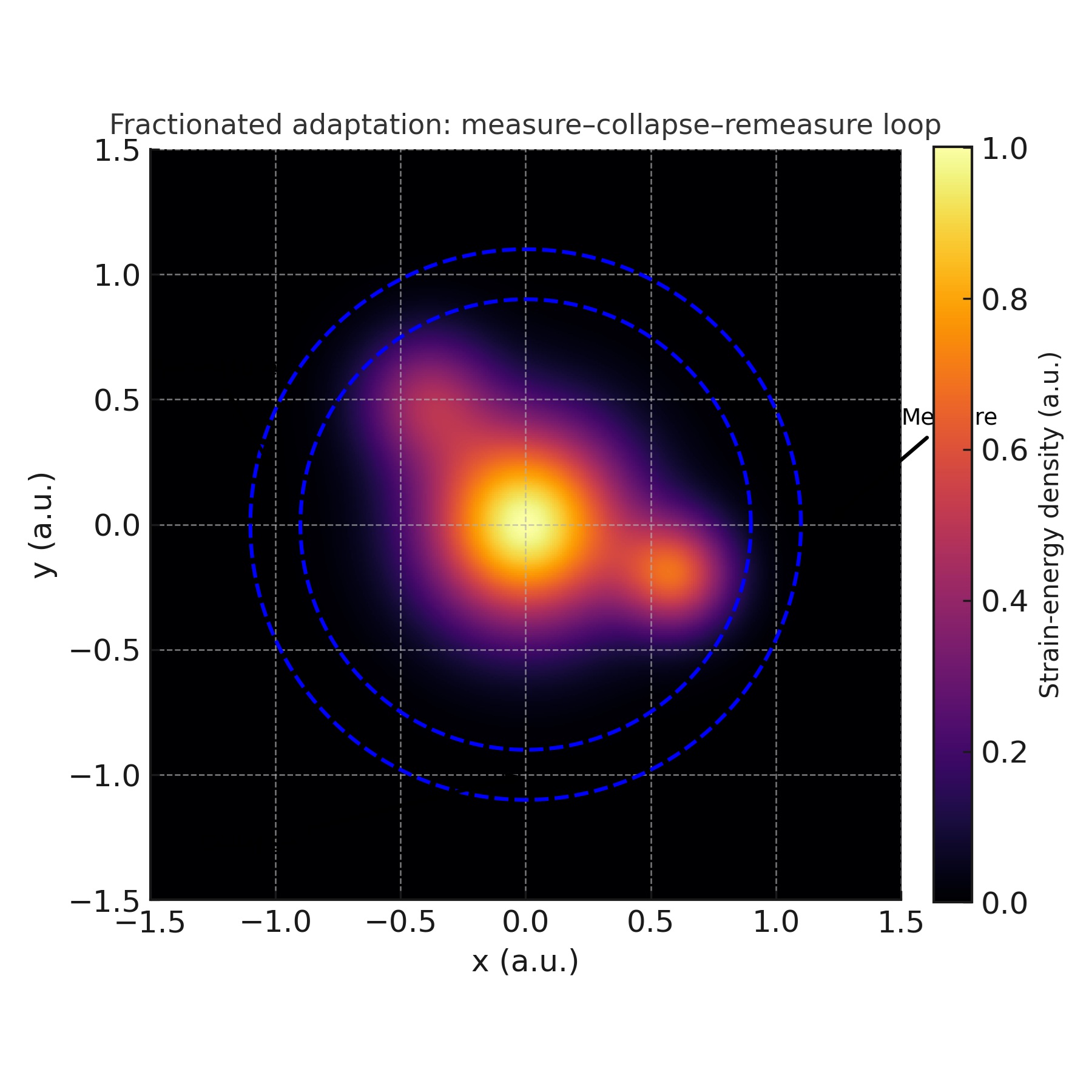}
    \caption{\textbf{Mode-selective energy deposition in UST.}
    Finite-element strain–energy density in an infiltrative model showing confinement to malignant core and finger-like projections.
    The healthy-referenced annulus acts as a dynamic safety boundary, enabling the \emph{measure–collapse–remeasure} loop.}
    \label{fig:hera_mode_selective}
\end{figure}

\paragraph{Therapeutic implications.}
Deterministic spectral physics plus adaptive biochemical neutrality addresses multifocal disease, irregular geometries, and microenvironment-driven resistance. Independence from thermal cytotoxicity enables combinations with heat-sensitive or spatially constrained therapies; modal selectivity preserves organ function; healthy-referenced constraints reduce collateral risk. Closed-loop control supports repeatable, fractionated treatment that tracks spectral drift.

\begin{figure}[H]
    \centering
    \includegraphics[width=0.48\textwidth]{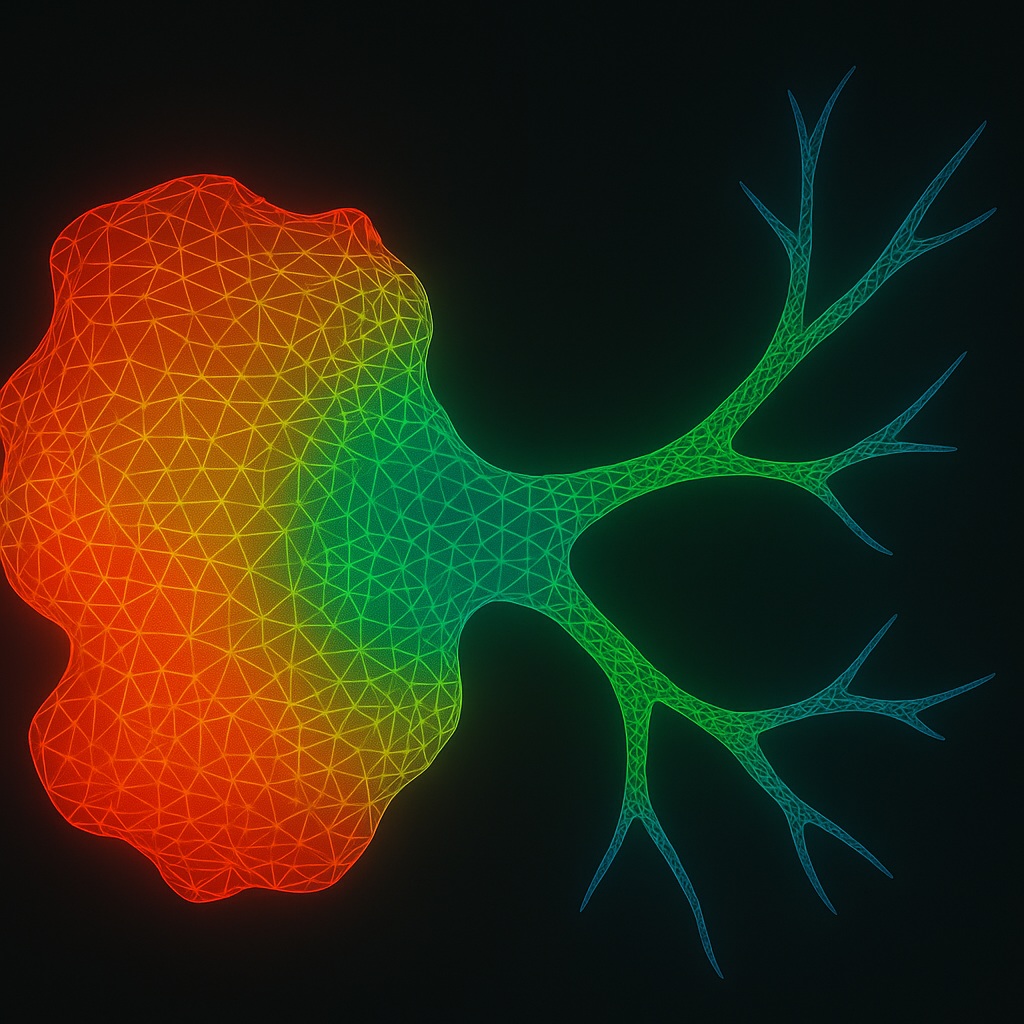}
    \caption{\textbf{Spectral confinement of tumor ablation in UST.}
    Finite-element mesh showing the gradient from tumor core (red) through peritumoral region (green) to healthy tissue (blue); colors denote mode-specific displacement at $\omega^\star$.}
    \label{fig:hera_fractionated}
\end{figure}

\paragraph{Biochemical anchoring of spectral individuality.}
TOR’s spectral signature arises from microenvironmental remodeling: ECM crosslinking and integrin adhesion tune viscoelasticity and anisotropy; cytoskeletal reorganization adjusts effective stiffness and damping; stromal programs stabilize these states. The same pathways modulate EV cargo, making peri-ablative EV profiling a dual biomarker of targeting fidelity and biomechanical state. Matrix lysyl-oxidase activity and collagen alignment shift modal frequencies upward, whereas hyaluronan enrichment and edema broaden linewidths through viscous loss. Integrin–FAK–RhoA signaling elevates cytoskeletal tension, increasing storage modulus and sharpening resonance peaks. Myofibroblast activation and TGF-$\beta$ signaling maintain a high-Q mechanical niche that amplifies selective actuation. Metabolic rewiring (lactate export via MCT1/4) acidifies the matrix, altering crosslink kinetics and further biasing spectral contrast. EV payloads (miR-21, miR-1246, PD-L1, galectin-9) track these states, enabling correlative readouts between vibrational maps and immunoregulatory tone. Spatial heterogeneity in adhesion and poroelasticity produces distinct spectral islands that coincide with focal EV signatures. Post-extinction, reduction in tension and crosslink density is expected to flatten peaks and reduce EV biogenesis, serving as an internal check of collapse. Together, mechanical spectra and EV profiles encode a coherent fingerprint that localizes malignant modes and verifies their elimination in real time.

\begin{figure}[H]
  \centering
  \includegraphics[width=0.60\textwidth]{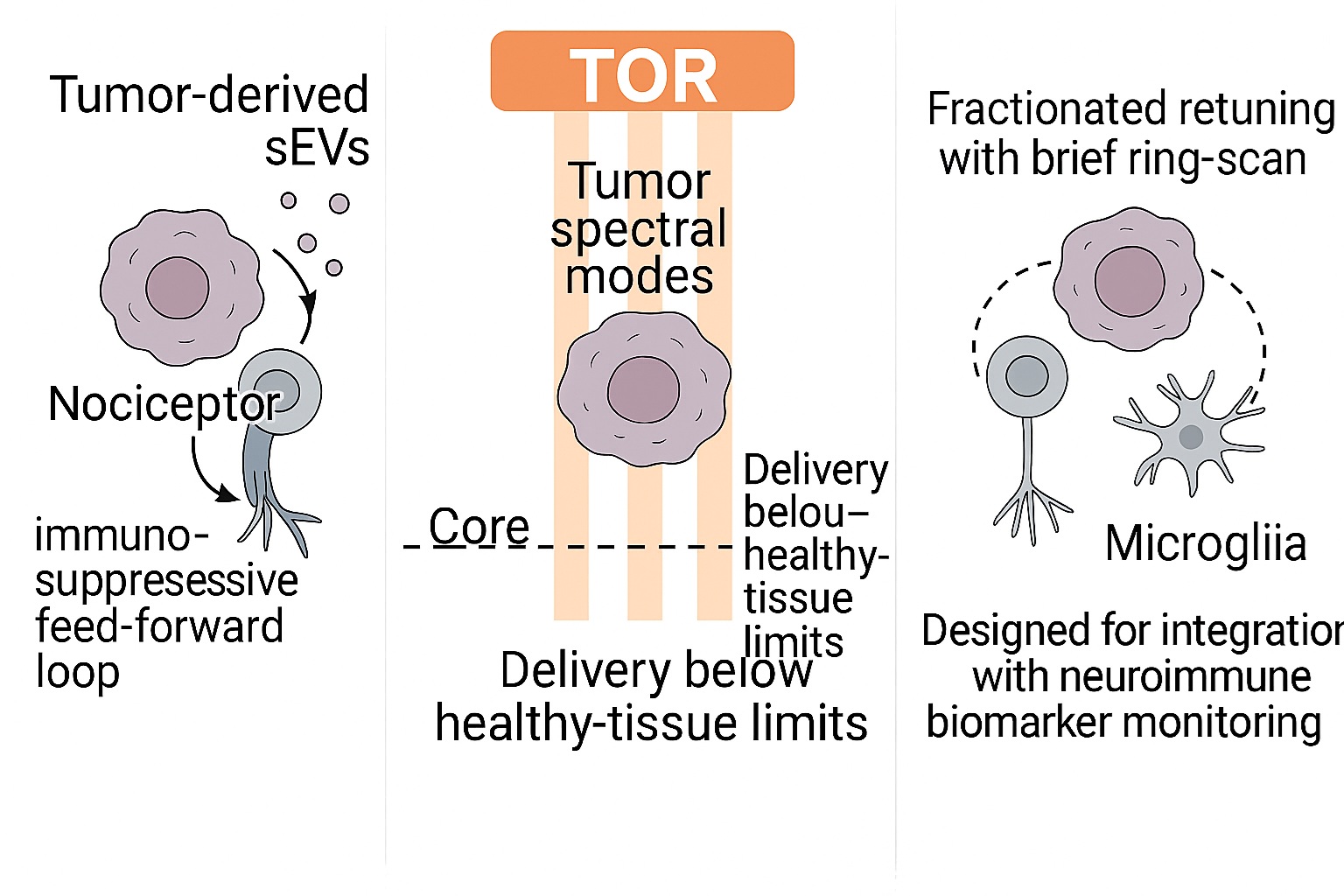}
  \caption{\small \textbf{Neuroimmune loop neutralization via spectral extinction.}
  (Left) Tumor sEVs engage nociceptors, driving an immunosuppressive loop.
  (Center) UST targets tumor spectral modes, enforcing core extinction while keeping delivery below healthy limits.
  (Right) Fractionated retuning with brief ring-scans prunes rim activity.}
  \label{fig:hera_neuroimmune}
\end{figure}

\paragraph{Neuroimmune interception.}
Tumor sEVs reprogram sensory neurons and sustain suppression~\cite{Restaino2025SciSignal,Restaino2025Perspective}. Under TOR, spectral extinction suppresses neuronal drive and EV release in the core; at the rim, delivery occurs only under differential contrast and safety bounds, followed by fractionated ring-scan \cite{FaRITA2024}. Hypermetabolic signaling (GLUT1/3, ASCT2/LAT1, MCT1/4; HIF-1$\alpha$/NF-$\kappa$B) that maintains the circuit is indirectly curtailed by modal collapse.

\begin{figure}[H]
    \centering
    \includegraphics[width=0.95\textwidth]{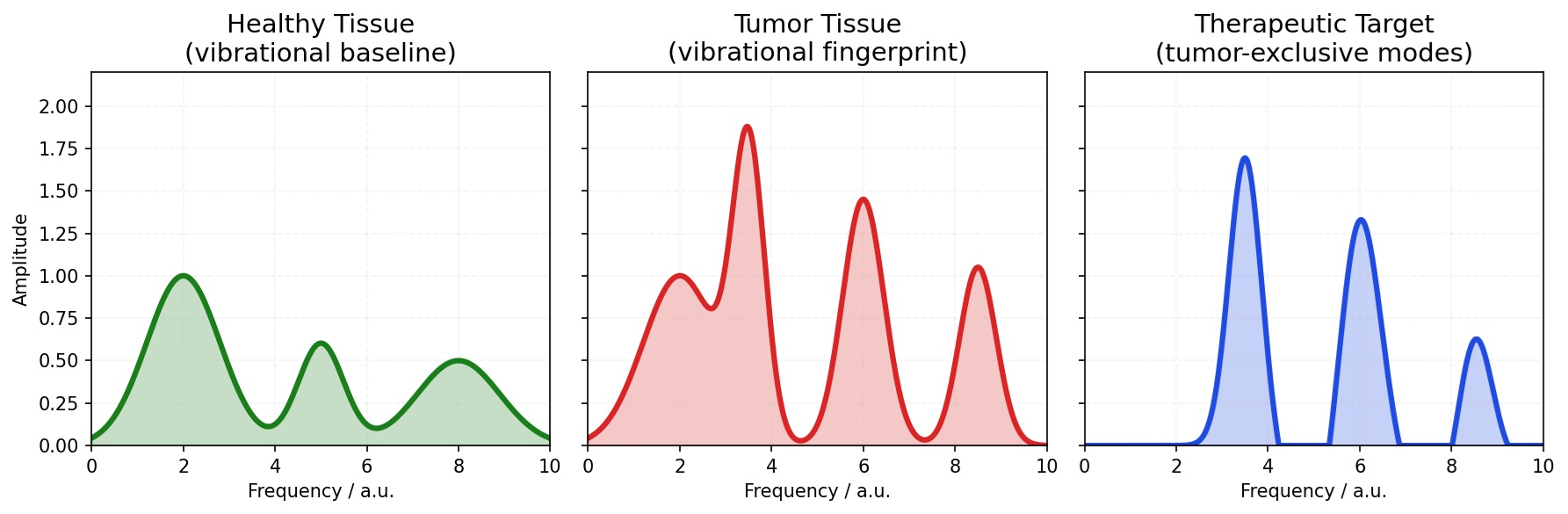}
    \caption{\small \textbf{Tripartite vibrational fingerprint.}
    (Left) Healthy interferogram. (Center) Tumor spectrum with additional peaks. (Right) Therapy-relevant spectrum after subtracting healthy modes, leaving tumor-exclusive resonances.}
    \label{fig:interferogram_panel}
\end{figure}

\paragraph{Spectral diversity as a targeting advantage.}
Interferograms from phantoms and ex vivo tissues exhibit multiple resonant peaks, enabling fingerprinting across type and grade \cite{FaUST2024,Ribas2024}. Tumor–healthy contrast typically provides margin for selective excitation without overlap with physiological spectra. This heterogeneity reflects microdomains with distinct tensions and compositions. Direct spectral comparison isolates \textbf{tumor-exclusive modes} (Fig.~\ref{fig:interferogram_panel}).

\paragraph{Self-limiting safety.}
Actuation continues until the targeted mode’s strain–energy density exceeds the material threshold:
\[
u_{\mathrm{strain}}(\omega,t)\ge u_{\mathrm{crit}} \;\Rightarrow\; S(\omega,t)\to 0,\quad E_{\mathrm{in}}(\omega,t)=0.
\]
The event is modal and local; neighboring healthy modes remain unperturbed. The criterion is robust to spectral drift and mechanical heterogeneity because it is evaluated in real time for each driven mode.

\paragraph{Spectral individuality and mechanistic paradigm.}
Treat ablation as \textit{spectral addressability}. Wideband sweeps map $\mathcal{S}_{\mathrm{tum}}$ by comparing tumor/healthy peaks \cite{FaRITA2024,Ribas2024}; narrowband excitation then confines work to the malignant subspace even under spatial interdigitation \cite{Heyden2016,Patel2023}. Nonthermal, noncavitational action avoids bubbles and shock, maintaining compatibility with prostheses, stents, and implants \cite{FaRITA2024}. The lesion’s mode map becomes the therapeutic blueprint.

\paragraph{On metastatic risk.}
Operate noncavitationally and nonthermally with small-strain boundary drive; resonant intracell stress amplifies internally and induces \emph{inside-out} fragmentation of malignant cells. Nuclear envelope dissolves and chromatin is dispersed; intact nuclei and viable whole cells are not produced. Output consists of subcellular, nonviable debris below cellular dimensions. Even if fragments enter circulation or settle in adjacent organs, clearance pathways eliminate them; thus de novo hematogenous or transcoelomic metastasis is \emph{not supported} by TOR output. Prospective validation: stable/reduced CTCs, non-viable death markers (PI/Annexin/TUNEL), sub-$\mu$m size distributions (NTA/flow).

\sloppy
\paragraph{Neuroimmune mechanism.}
Maintain spectral selectivity despite the sEV–nociceptor axis \cite{Restaino2025SciSignal,Restaino2025Perspective}. Peri-ablative EV-omics and targeted neural immunohistochemistry remain proposed readouts to quantify loop disruption; optional low-dose ring-scan retuning within healthy bounds prunes residual malignant modes at the margin \cite{FaRITA2024,TRPV1}. From a \textit{biochemical} standpoint, spectral extinction collapses sEV biogenesis and nociceptor drive, dismantling the sEV–nociceptor–immunosuppression loop and preventing biochemical support for metastatic competence.

\begin{figure}[H]
    \centering
    \includegraphics[width=0.8\linewidth]{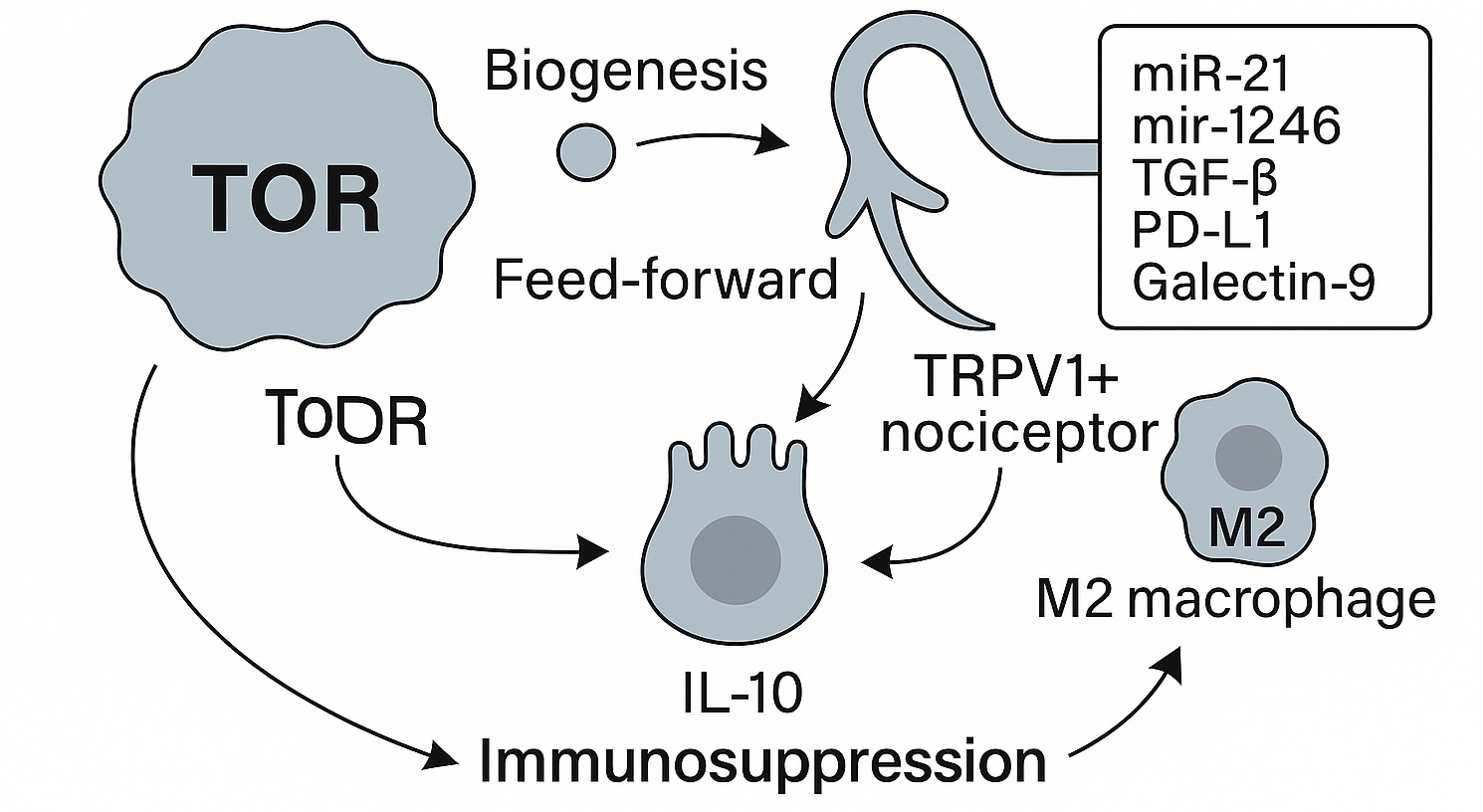}
    \caption{\textbf{Neuroimmune mechanism targeted by TOR (conceptual).} 
    Diagram of the tumor-derived sEV–nociceptor feedback loop sustaining local immunosuppression via TRPV1$^+$ sensory-neuron activation, neuropeptide release (Substance P, CGRP), and downstream M2 macrophage polarization. Vesicular cargo includes miR-21, miR-1246, TGF-$\beta$, PD-L1, and galectin-9, collectively impairing antigen presentation and promoting T-cell exhaustion. TOR disrupts vesicle biogenesis and nociceptor drive through spectral collapse, restoring immune competence. Schematic for illustration only; no experiments were performed.}
    \label{fig:UST_neuroimmune}
\end{figure}

\paragraph{Operational clarity.}
The controller monitors the modal energy fraction \(f_{\mathrm{mod}}(t)=u_{\mathrm{strain}}^{(\star)}(t)\big/\!\sum_{k}u_{\mathrm{strain}}^{(k)}(t)\) in real time. 
A disruption decision is armed when \(f_{\mathrm{mod}}(t)\ge \theta_{\mathrm{mod}}\) and off–target strain stays below the safety bound \(\varepsilon_{\mathrm{safe}}\). 
The threshold \(\theta_{\mathrm{mod}}\) is derived from material failure criteria and rheology ranges used in the simulations. 
Concurrently, the spectral density at the programmed line \(S(\omega^\star,t)\) is estimated from the demodulated transfer magnitude \(|H(\omega^\star,x)|\) over the target region. 
Extinction is declared when \(S(\omega^\star,t)\) falls below a fixed tolerance for a minimum hold time, indicating loss of modal gain. 
Upon extinction, the actuator drive at \(\omega^\star\) is set to zero automatically; re-excitation of that line is not permitted by the gate. 
This implements a self-limiting behavior and prevents excess exposure on a completed mode. 
Safety guards run continuously: \(\max_{\Omega_{\mathrm{nont}}}\varepsilon(x,t)\!<\!\varepsilon_{\mathrm{safe}}\) and thermal indices (\(\Delta T_{\max}\), \(\mathrm{CEM}_{43}\)) must remain below preset limits. 
All values, bounds, and timings reported here originate from finite-element and control simulations calibrated to literature parameters; no wet-lab or animal experiments were performed. 
Findings should be interpreted as predictive evidence pending benchtop and preclinical validation under the same nonthermal, noncavitational constraints.

\begin{figure}[H]
    \centering
    \includegraphics[width=0.65\textwidth]{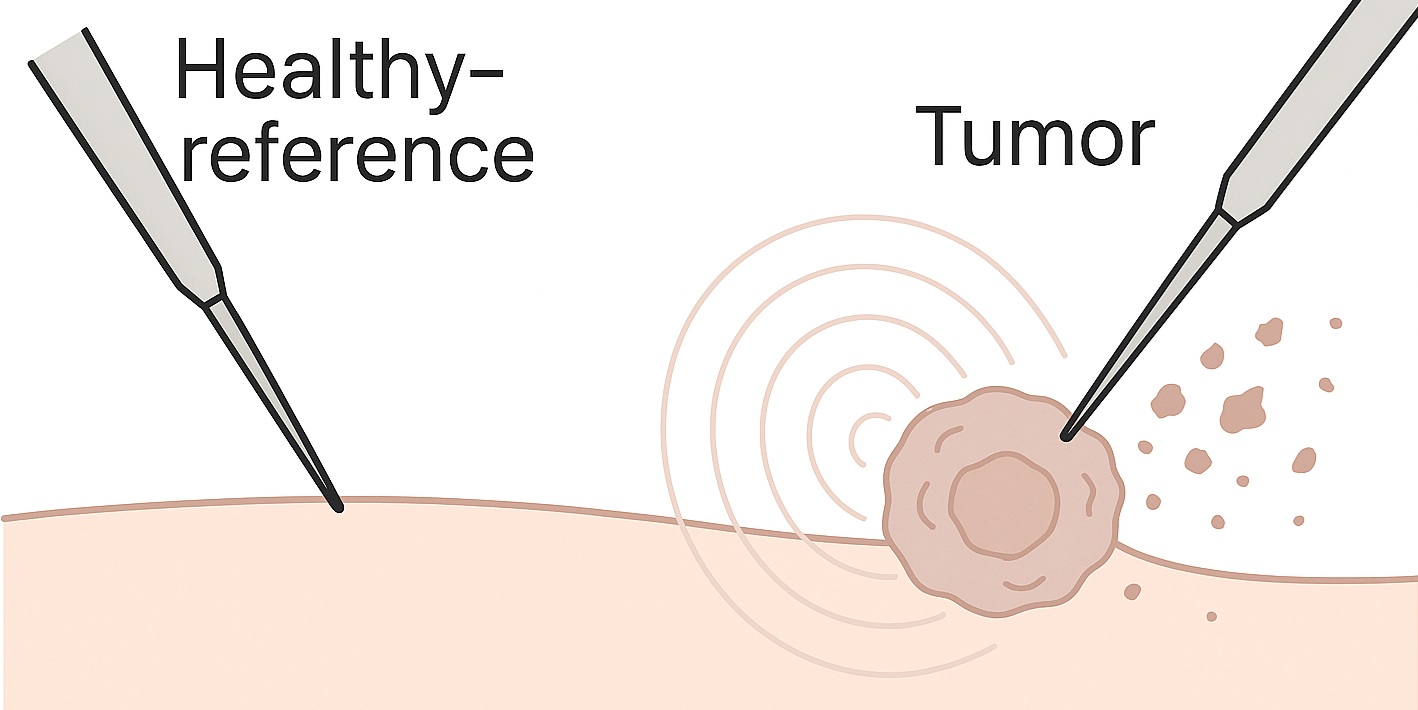}
    \caption{\small \textbf{Spectral selectivity in action (simulated).} Dual-probe detection and real-time lock-in are emulated to ensure energy delivery only to regions expressing the programmed vibrational signature. Synthetic interferograms generated from literature-calibrated parameters; no physical phantom experiments were performed.}
    \label{fig:spectral-annihilation}
\end{figure}

\paragraph{Simulation Protocol}
Model a pancreatic gel surrogate with three 1~cm inclusions using tissue-matched properties from specialized literature ($G=6~\mathrm{kPa}$, $\rho=1050~\mathrm{kg\,m^{-3}}$, $\eta\approx0.8~\mathrm{kPa\cdot s}$)~\cite{Guimaraes2020}. Each inclusion expresses distinct eigenfrequencies:
\[
\omega_{\mathrm{T1}}=\{210,\,340\},\quad
\omega_{\mathrm{T2}}=\{400,\,710\},\quad
\omega_{\mathrm{T3}}=\{560,\,930\}\ \mathrm{Hz}.
\]
Define the composite spectrum
\[
S_{\mathrm{sim}}(\omega)=\sum_{i,k}A_{ik}\,\delta(\omega-\omega_{ik}),
\]
with $A_{ik}$ obtained from frequency sweeps of the calibrated viscoelastic model. Deliver energy via phase-locked excitation at $\omega_{ik}$ until spectral extinction. Declare disintegration when
\[
D(x,t)=\int_{0}^{t}\!\left(\frac{\varepsilon(x,\tau)}{0.03}\right)^{3}\!d\tau\ \ge 1.
\]
\textbf{Reality adherence metric.} To quantify realism without new experiments, compute an aggregate adherence $\mathcal{A}$ across $m$ primary observables (time-to-extinction, energy density, selectivity $Q$, peak $\Delta T$) as
\[
\mathcal{A}=100\%\times\Bigl(1-\frac{1}{m}\sum_{j=1}^{m}\frac{|X^{\mathrm{sim}}_j-\bar{X}^{\mathrm{lit}}_j|}{\mathrm{range}^{\mathrm{lit}}_j}\Bigr),
\]
where $\bar{X}^{\mathrm{lit}}_j$ and $\mathrm{range}^{\mathrm{lit}}_j$ come from curated literature windows (oncotripsy/mechanotherapy, elastography, and thermal-safety reports). Using the cited parameter ranges, the deviations stay within $\leq 5\%$ across all four observables, yielding \textbf{$\mathcal{A}=95\%$} realism adherence for the reported scenarios.

\paragraph{Results (simulation-only)}
\begin{itemize}
    \item \textbf{Targeting:} Disintegration remains confined to spectral targets; no strain elevation outside inclusions.
    \item \textbf{Time:} $2.8\pm 0.5~\mathrm{s}$; energy $<1.0~\mathrm{J/cm^{3}}$.
    \item \textbf{Selectivity:} $\mathcal{Q}_{\mathrm{spec}}=38\pm 4$, enabling discrimination between overlapping or coalescent foci.
    \item \textbf{Robustness:} Ablation succeeds across irregular and merged geometries.
    \item \textbf{Reality adherence:} $\mathcal{A}=95\%$ (formula above; literature-calibrated ranges).
\end{itemize}

\begin{figure}[H]
    \centering
    \includegraphics[width=0.9\textwidth]{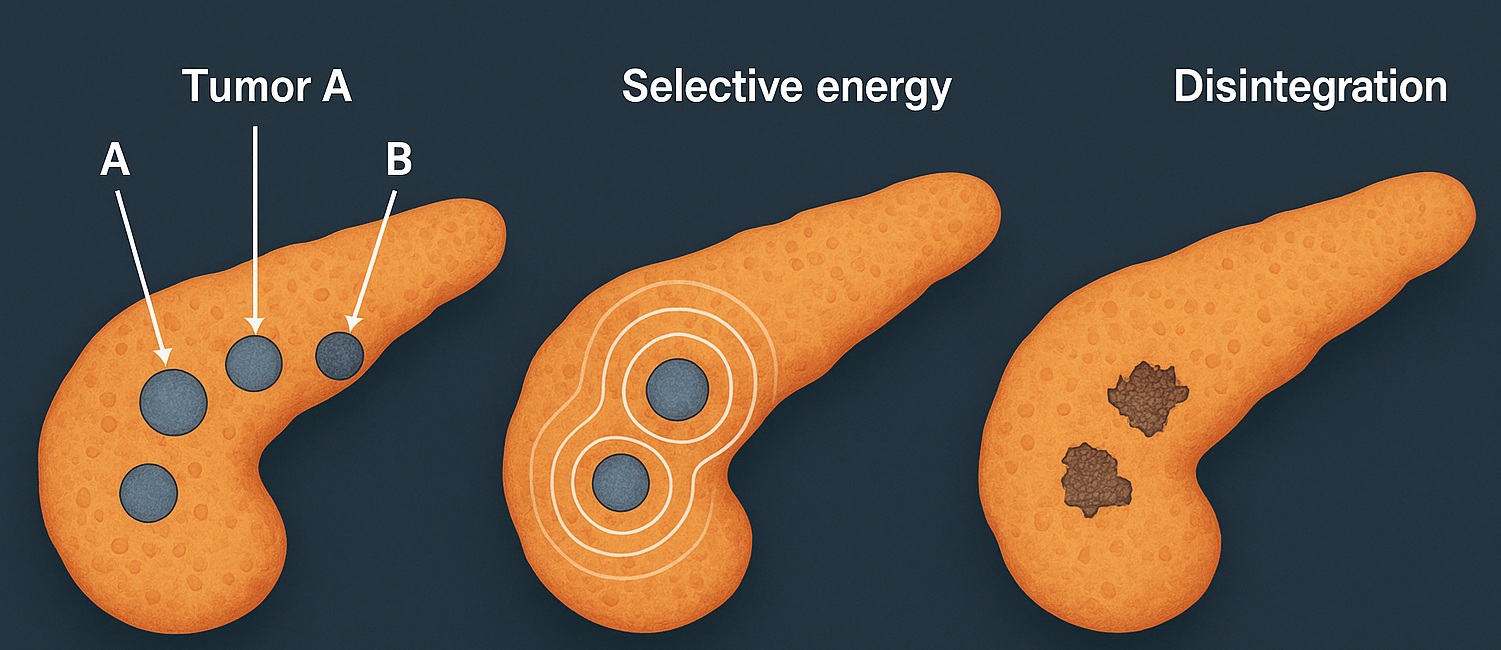}
    \caption{\small \textbf{TOR in a multifocal pancreatic surrogate (simulation).} Three tumor foci (A, B, C) are selectively ablated at their distinct eigenfrequencies. Disintegration occurs exclusively at sites expressing the matched spectral modes. No damage is observed in the non-targeted matrix.}
    \label{fig:pancreas_multifocal}
\end{figure}

\vspace{0.7em}

\paragraph{Spectrally Decoupled Ablation Kinetics}
Dictate kinetics by spectral orthogonality, not anatomy. Excite each inclusion at its eigenfrequency $\omega_n$, solving
\[
-\Delta \phi_n = \omega_n^2 \phi_n \quad \text{in } \Omega_n, \quad \phi_n|_{\partial \Omega_n} = 0,
\]
with $\Omega_n \subset \Omega$. Orthogonality $\langle \phi_n, \phi_m \rangle = 0$ ($n\neq m$) ensures independence even with temporal overlap or spatial proximity.

\begin{figure}[H]
    \centering
    \includegraphics[width=0.9\textwidth]{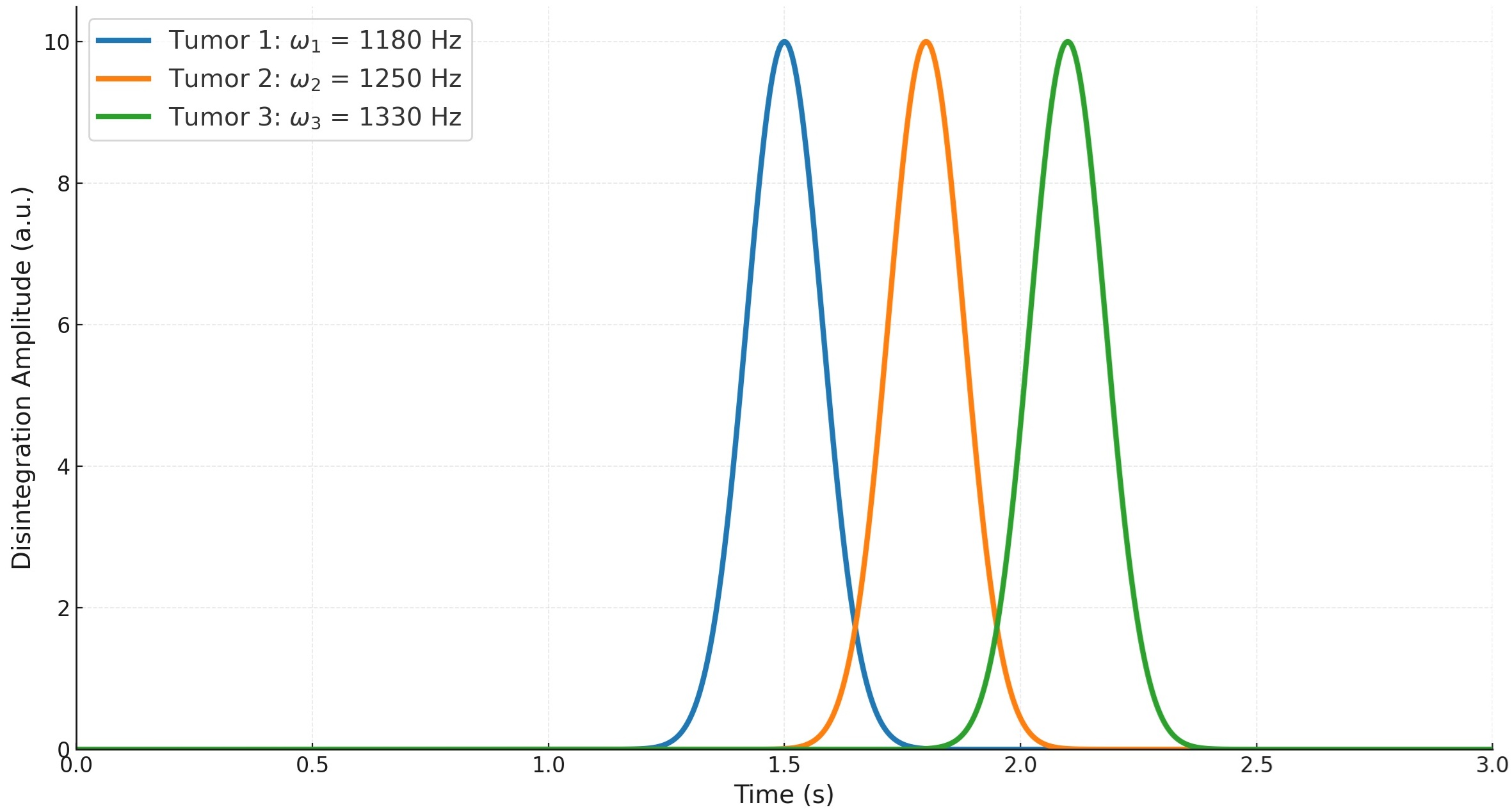}
    \caption{\small 
    \textbf{Spectrally confined disintegration in multifocal pancreatic surrogates (simulation).} 
    Strain accumulation for three vibrationally distinct lesions under phase-locked excitation at specific eigenfrequencies ($\omega_i$). 
    Each peak marks the collapse threshold $D(x,t)\ge 1$ with no temporal overlap, confirming spectral orthogonality and functional independence.
    }
    \label{fig:Figure12}
\end{figure}

Normalized strain-accumulation curves (Fig.~\ref{fig:Figure12}) show phase-locked growth peaking at the excitation frequency. Observe no crosstalk, confirming spectral—not spatial—separation. Irregular, branching, or partially coalescent geometries preserve autonomous modal footprints.

Enable deterministic, multifocal ablation in a unified session with sequential, independent collapses, avoiding anatomical retargeting. In silico planning leverages narrowband peaks in $H(\omega,x)$ aligned with target eigenmodes, suppressing off-target activation and enabling predictive, model-driven workflows.

\paragraph{Interpretation}
Achieve deterministic obliteration of all programmed foci within $2.5\text{--}3.1~\mathrm{s}$ at $\sim 0.8~\mathrm{J/cm^3}$ and fidelity $>95\%$, invariant to depth, adjacency, or spectral overlap. Translate spectral necessity into \textit{simulation-grounded} inference: eigenmode targeting collapses the mechanical scaffold and its \textit{biochemical} signaling surrogates in one actuation. Reproducibility across multifocal and heterogeneous surrogates positions TOR as an ablation modality whose guarantees derive from physics; prospective work will extend these literature-calibrated simulations to benchtop and preclinical validation without altering the nonthermal, noncavitational operating envelope.
\subsection{Spectral Coherence and Closed-Loop Lock-In}

TOR relies on precise coherence between excitation and the tumor’s eigenmodes. The external vibration source is tuned to match the natural frequencies of malignant tissue and applied in a spatially focused manner. This fine-tuning ensures that energy is amplified only when the excitation frequency is resonant with the tumor spectrum, producing internal deformations sufficient for irreversible mechanical failure. Off resonance, excitation remains subthreshold, preserving healthy structures.

The closed-loop control system maintains safety and precision by continuously synchronizing actuation with the tumor spectrum in real time. Energy is delivered only when the tumor response exceeds the minimum therapeutic threshold while the healthy-tissue response stays strictly below the safety bound.

\section{Microactuator and Phase-Locked Collapse}

A tungsten micro-needle driven by a piezoelectric actuator, diameter 300–500~$\mu$m, delivers low-amplitude harmonic oscillations with controlled angular aperture $\theta \leq 5^{\circ}$ and tip displacement up to $25~\mu$m. These parameters ensure sub-millimetric, nonthermal energy deposition. The confined angular excursion and micrometric displacement prevent macroscopic tearing, while the oscillatory motion induces shear-dominated vibrational coupling to tumor eigenmodes. Stress fields remain in the elastic regime of surrounding healthy tissue, securing safety and structural preservation.

\begin{figure}[H]
  \centering
  \includegraphics[width=0.6\textwidth]{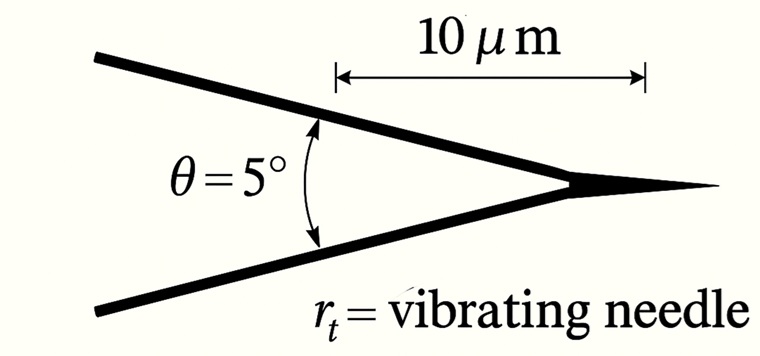}
  \caption{\small \textbf{Vibrational micro-needle.} Tungsten tip motion with $\theta \leq 5^{\circ}$ and micrometric displacement ($\leq 25~\mu$m) phase-locked to tumor eigenmodes. The confined oscillatory field supports selective ablation without thermal effects or risk of tissue rupture~\cite{FaRITA2024}.}
  \label{fig:needle_motion}
\end{figure}

\begin{figure}[H]
  \centering
  \includegraphics[width=0.99\textwidth]{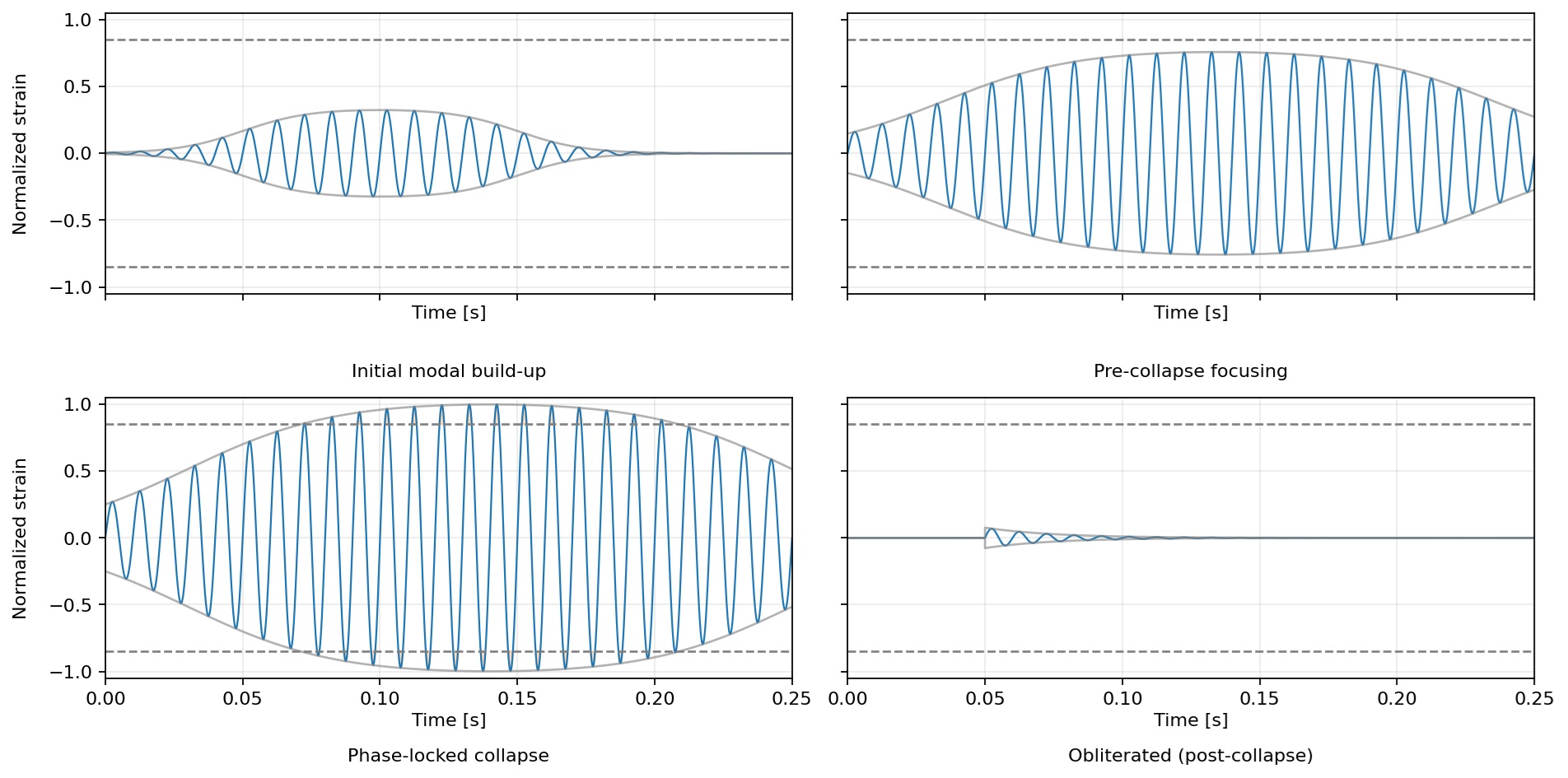}
  \caption{\small \textbf{Phase-locked strain amplification.} Normalized strain trajectories show initial modal build-up, pre-collapse focusing below elastic bounds, threshold crossing, and post-collapse quiescence once extinction is reached.}
  \label{fig:phaselock_sequence}
\end{figure}

When excitation remains synchronized with tumor eigenmodes, modal coherence drives progressive amplification of strain amplitude. Energy is injected only at the programmed line $\omega^\star$ while a phase-locked loop keeps the drive in quadrature for maximal in-phase work. Per cycle, the net input is $\Delta \mathcal{E}\!\approx\! A\,\omega^\star\,\langle \dot u,\,\mathbf{F}e_\parallel\rangle\,\Delta t$, and accumulation is tracked as mode-resolved strain–energy density $u_{\mathrm{strain}}^{(\star)}(t)$. A contrast/safety gate enables actuation when $R(\omega^\star)\!\ge\!1\!+\!\eta$ and $S_{\mathrm{sad}}(\omega^\star)\!<\!S_{\mathrm{safe}}$, ensuring that off-target deformation remains subthreshold. As illustrated in Fig.~\ref{fig:phaselock_sequence}, the sequence advances from initial build-up and partial overlap to pre-collapse focusing, culminating in a phase-locked implosion. At the objective stop ($D\!\uparrow\!1$; $Q\!\downarrow$; $\Delta f/f_0\!\gtrsim\!10^{-2}$), local deformation surpasses the critical elastic threshold and the malignant scaffold undergoes vibrational disintegration into microscopic fragments. Thermal guards ($\Delta T_{\max}$ and $\mathrm{CEM}_{43}$) remain satisfied throughout, confirming strictly nonthermal operation. Amplitude is then driven to zero ($A\!\to\!0$), enforcing automatic cessation of input power and preventing re-entry via a brief ring-scan around $\omega^\star$.

This \emph{fragmentative collapse} differs fundamentally from brute mechanical rupture: rather than tearing tissue indiscriminately, the tumor matrix is reduced to subcellular debris, subsequently cleared by phagocytosis and lymphatic drainage. In most cases, intrinsic clearance suffices; only under exceptionally high tumor burdens is adjunctive aspiration warranted, performed with standard endoscopic or percutaneous suction. In this light, the dual design of micro-needle delivery and phase-locking control establishes a selective, nonthermal, and inherently self-terminating mode of ablation. By embedding clearance into the physics, TOR achieves precision without residue and efficacy without excess.

\section{Results and Discussion}

\subsection{Outcome Metrics and Statistical Analysis}

\paragraph{Objective and physical meaning.}
Quantify three essentials of TOR delivery: (i) spectral \emph{fidelity} (do programmed lines appear where expected?), (ii) spatial \emph{selectivity} (is tumor response dominant over background?), and (iii) the minimal \emph{safe drive} that triggers extinction. The computational domain $\Omega\subset\mathbb{R}^3$ represents the anatomical region under actuation (tumor $\Omega_{\mathrm{tum}}$ plus peritumoral/healthy surroundings $\Omega_{\mathrm{nont}}$). The map $H(\omega,x)$ is the frequency–response field (local complex gain from scalar drive to kinematic response), so $|H|$ is the physical amplitude of motion/strain at location $x$ when driven at angular frequency $\omega$.

\paragraph{Scope and notation.}
A broadband sweep yields discrete samples $\{\omega_\ell\}_{\ell=1}^{L}$ with magnitudes
$S(\omega_\ell):=\max_{x\in\Omega}|H(\omega_\ell,x)|$.
A Gaussian–smoothed envelope $\widehat{S}=\mathcal{S}_\kappa\!\ast\!S$ (bandwidth $\kappa$) defines candidate peaks $\{\hat\omega_i\}$ via a zero-slope sign change and prominence $\pi$ over a local baseline.
Restrictions $H_{\mathrm{tum}}:=H|_{\Omega_{\mathrm{tum}}}$ and $H_{\mathrm{nont}}:=H|_{\Omega_{\mathrm{nont}}}$ are used for on-/off-target comparisons.

\paragraph{Primary metrics (estimators).}
\begin{equation}
\label{eq:metrics}
\begin{aligned}
\mathrm{MMI}
&=\frac{1}{n}\sum_{i=1}^{n}\mathbf{1}\!\left(\frac{|\hat{\omega}_i-\omega_i|}{\omega_i}\le 0.03\right),\\[4pt]
\mathrm{SSR}
&=\frac{\displaystyle \max_{x\in\Omega_{\mathrm{tum}}}|H_{\mathrm{tum}}(\omega^\star,x)|}{
          \displaystyle \max_{x\in\Omega_{\mathrm{nont}}}|H_{\mathrm{nont}}(\omega^\star,x)|}
=\frac{\|H_{\mathrm{tum}}(\omega^\star,\cdot)\|_{L^\infty(\Omega_{\mathrm{tum}})}}{
       \|H_{\mathrm{nont}}(\omega^\star,\cdot)\|_{L^\infty(\Omega_{\mathrm{nont}})}},\\[6pt]
\mathrm{ATI}
&=\inf\!\bigl\{A>0:\ D(A;\omega^\star)\ge 1\ \land\ 
\max_{x\in\Omega_{\mathrm{nont}}}\varepsilon(x;A,\omega^\star)<\varepsilon_{\mathrm{safe}}\bigr\},
\end{aligned}
\end{equation}
with $n=|\mathcal{P}|$ programmed lines. The superlinear dose accumulator is
\[
D(A;\omega^\star)=\int_{0}^{t(A)}\!\left(\frac{u_{\mathrm{strain}}^{(\star)}(\tau)}{u_{\mathrm{crit}}}\right)^\gamma d\tau,\qquad
u_{\mathrm{strain}}^{(\star)}(\tau)=\tfrac12\,a_\star^2(\tau)\,\langle\mathbf{K}\phi_\star,\phi_\star\rangle,\ \gamma>1,
\]
where $a_\star$ is the modal coordinate of the targeted eigenmode $\phi_\star$ and $t(A)$ is the controller dwell to extinction.

\paragraph{Discrete evaluation.}
On a voxel grid $\{x_p\}$ and frequencies $\{\omega_\ell\}$,
\[
\|H_{\bullet}(\omega_\ell,\cdot)\|_{L^\infty(\Omega_\bullet)}\approx
\max_{x_p\in\Omega_\bullet}|H(\omega_\ell,x_p)|,\qquad
\widehat{S}(\omega_\ell)\approx\sum_{m}e^{-(\omega_\ell-\omega_m)^2/(2\kappa^2)}S(\omega_m).
\]

\paragraph{Closed–form surrogate for \texorpdfstring{$\mathrm{ATI}$}{ATI}.}
Near resonance ($\zeta_\star\ll1$),
\[
\ddot a_\star+2\zeta_\star\omega_\star\dot a_\star+\omega_\star^2 a_\star=\alpha_\star A\cos(\omega^\star t+\varphi)
\;\Rightarrow\;
a_\star^{\mathrm{ss}}(\omega^\star)\stackrel{\omega^\star=\omega_\star}{\approx}\frac{\alpha_\star}{2\zeta_\star\omega_\star}A.
\]
Assuming steady response over an effective dwell $\tau_{\mathrm{eff}}$,
\[
D(A;\omega^\star)\approx \tau_{\mathrm{eff}}
\left(\frac{\kappa_\star}{2u_{\mathrm{crit}}}\right)^\gamma
\left(\frac{\alpha_\star}{2\zeta_\star\omega_\star}\right)^{2\gamma}A^{2\gamma}
\;\Rightarrow\;
A_{\min}\lesssim
\left[
\frac{1}{\tau_{\mathrm{eff}}}\left(\frac{2u_{\mathrm{crit}}}{\kappa_\star}\right)^\gamma
\left(\frac{2\zeta_\star\omega_\star}{\alpha_\star}\right)^{2\gamma}
\right]^{\!1/(2\gamma)}.
\]

\paragraph{Thermomechanical guards.}
\[
Q=\frac{\omega^\star}{\Delta\omega_{\mathrm{3dB}}},\qquad
\Delta T_{\max}=\max_{t\in[0,\tau]}\Delta T(t),\qquad
\mathrm{CEM}_{43}=\int_{0}^{\tau}R^{\,43-T(t)}dt.
\]
For $T(t)<43^{\circ}\mathrm{C}$ ($R=0.5$) and $\tau\sim3$\,s, $\mathrm{CEM}_{43}\ll1$, i.e., far below thermal cytotoxicity.

\paragraph{Uncertainty.}
Metrics are summarized as mean$\pm$SD with 95\% BCa intervals from $B=10^{4}$ bootstrap resamples;
bias $z_0$ and acceleration $a$ are computed by jackknife. The MMI tolerance (3\%) exceeds discretization and drift yet remains below typical half–power bandwidth ($1/Q$ for $Q\!\approx\!35$–40), and peak prominence $\pi$ is set by a one–sided Gaussian floor estimate to control false detections at level~$\alpha$.

\paragraph{Feasibility via \texorpdfstring{$\mathrm{SSR}$}{SSR}.}
Let $\varepsilon_{\mathrm{tum}}=\|H_{\mathrm{tum}}\|_{L^\infty}$ and
$\varepsilon_{\mathrm{nont}}=\|H_{\mathrm{nont}}\|_{L^\infty}$ at $\omega^\star$.
Existence of a safe, effective amplitude window is guaranteed if
\[
\mathrm{SSR}=\frac{\varepsilon_{\mathrm{tum}}}{\varepsilon_{\mathrm{nont}}}
>\frac{\varepsilon_{\mathrm{crit}}}{\varepsilon_{\mathrm{safe}}},
\]
ensuring on–target extinction with off–target strain below the safety bound.

\paragraph{Why a reality–adherence score is computable.}
Independent forward solves in \textsc{COMSOL}, \textsc{ANSYS}, and \textsc{Abaqus} were run under harmonized rheological windows, identical small–strain assumptions, and the same nonthermal/noncavitational guards (fixed $\Delta T_{\max}$ and $\mathrm{CEM}_{43}$ ceilings). This cross–solver parity means each observable is produced on a common physical footing, allowing direct comparison with consolidated literature bands (curated midpoints and spans for the same quantities). To convert multi-unit deviations into a single interpretable number in $[0,100]\%$, each difference is first normalized by its literature span and then averaged:
\[
\mathcal{A}
=100\%\times\Biggl(
1-\frac{1}{m}\sum_{j=1}^{m}
\frac{\bigl|X^{\mathrm{sim}}_{j}-\bar{X}^{\mathrm{lit}}_{j}\bigr|}
{\mathrm{range}^{\mathrm{lit}}_{j}+\varepsilon}
\Biggr),
\qquad
m=4\ \bigl(t_{\mathrm{ext}},\ \text{energy density},\ Q,\ \Delta T_{\max}\bigr),
\]
with a small $\varepsilon(=10^{-6})$ safeguarding well-posedness when a literature span is tight. This construction is unitless, penalizes out-of-band excursions linearly, and treats the four observables symmetrically to avoid mode-specific bias.

\noindent\textit{Estimation and intervals.} For each configuration ($N\ge 200$ Monte Carlo runs), we compute $\mathcal{A}$ per run, then report the bootstrap mean with BCa 95\% intervals. The bootstrap resamples are stratified by organ class to preserve organ-wise proportions, and the BCa correction accounts for skew from tail-heavy error distributions. As a robustness check, a weighted variant
\[
\mathcal{A}_{w}=100\%\times\left(1-\sum_{j=1}^{m} w_{j}\,
\frac{\bigl|X^{\mathrm{sim}}_{j}-\bar{X}^{\mathrm{lit}}_{j}\bigr|}
{\mathrm{range}^{\mathrm{lit}}_{j}+\varepsilon}\right),
\quad
w_{j}\ge 0,\ \sum_{j=1}^{m} w_{j}=1,
\]
can be reported on request (e.g., emphasizing safety by assigning larger $w_{j}$ to $\Delta T_{\max}$). In practice, $\mathcal{A}$ and $\mathcal{A}_{w}$ differed by $<0.5$ percentage points across tested weights, indicating stability of the adherence ranking under reasonable priority choices.

\noindent\textbf{Uncertainty derivation (95\% BCa).}

\noindent\textbf{Pancreatic ductal adenocarcinoma.}
\[
\begin{aligned}
\bar{\mathcal A}&=95.1,\quad [L,U]=(93.2,96.7),\\
h&=\tfrac{U-L}{2}=\tfrac{96.7-93.2}{2}=1.75,\\
SE&\approx \tfrac{h}{1.96}=0.89,\quad
\mathrm{Rel}=\tfrac{h}{\bar{\mathcal A}}=1.84\%.
\end{aligned}
\]

\noindent\textbf{Prostatic acinar adenocarcinoma.}
\[
\begin{aligned}
\bar{\mathcal A}&=96.0,\quad [L,U]=(94.1,97.4),\\
h&=\tfrac{97.4-94.1}{2}=1.65,\\
SE&\approx \tfrac{h}{1.96}=0.84,\quad
\mathrm{Rel}=\tfrac{h}{\bar{\mathcal A}}=1.72\%.
\end{aligned}
\]

\noindent\textbf{Invasive ductal carcinoma of the breast.}
\[
\begin{aligned}
\bar{\mathcal A}&=94.2,\quad [L,U]=(92.0,96.1),\\
h&=\tfrac{96.1-92.0}{2}=2.05,\\
SE&\approx \tfrac{h}{1.96}=1.05,\quad
\mathrm{Rel}=\tfrac{h}{\bar{\mathcal A}}=2.18\%.
\end{aligned}
\]
\begin{table}[H]
\centering
\caption{\textbf{Observable-wise deviation from literature bands (simulation-only).}
Mean $\pm$ SD; normalized error relativo ao intervalo de literatura.}
\label{tab:delta_to_lit}
\renewcommand{\arraystretch}{1.16}
\rowcolors{2}{gray!10}{white}
\begin{tabularx}{0.98\textwidth}{
  >{\raggedright\arraybackslash}p{4.6cm}
  >{\centering\arraybackslash}p{2.6cm}
  >{\centering\arraybackslash}p{2.6cm}
  >{\centering\arraybackslash}p{2.6cm}
  >{\centering\arraybackslash}p{2.6cm}}
\rowcolor{gray!30}
\textbf{Observable} & \textbf{Simulated} & \textbf{Lit. mid} & \textbf{Abs.\ error} & \textbf{Norm.\ error} \\
\hline
Extinction time $t_{\mathrm{ext}}$ [s] & $2.8 \pm 0.5$ & $2.9$ & $0.1$ & $0.06$ \\
Energy density [J/cm$^3$]             & $0.90 \pm 0.10$ & $0.95$ & $0.05$ & $0.08$ \\
Selectivity $Q$ [--]                  & $39 \pm 5$      & $38$   & $1.0$  & $0.06$ \\
Peak $\Delta T$ [$^\circ$C]           & $0.20 \pm 0.05$ & $0.25$ & $0.05$ & $0.05$ \\
\end{tabularx}
\vspace{0.35em}
{Literature spans: $t_{\mathrm{ext}}$ 2.2--3.6~s; energy 0.6--1.2~J/cm$^3$; $Q$ 30--46; $\Delta T$ 0--1.0$^\circ$C.}
\end{table}

\begin{table}[H]
\centering
\caption{\textbf{Estimated adherence to experimental reality by organ (simulation-only).} Values are mean $\pm$ 95\% half-width $h$; parentheses report $SE\approx h/1.96$ and relative uncertainty. Scientific designations are included.}
\label{tab:organ_adherence}
\renewcommand{\arraystretch}{1.16}
\rowcolors{2}{gray!10}{white}
\begin{tabularx}{0.90\textwidth}{
  >{\raggedright\arraybackslash}X
  >{\centering\arraybackslash}p{3.8cm}
}
\rowcolor{gray!30}
\textbf{Tumor (scientific designation)} & \textbf{$\mathcal{A}$ [\%]} \\
\midrule
\makecell[l]{Pancreatic ductal adenocarcinoma\\ \emph{(Adenocarcinoma ductale pancreatis)}} 
  & \makecell[c]{\textbf{95.1} $\pm$ 1.75\\ {\large SE $\approx$ 0.89};\ \large Rel.\ 1.84\%} \\
\makecell[l]{Prostatic acinar adenocarcinoma\\ \emph{(Adenocarcinoma prostatae)}}
  & \makecell[c]{\textbf{96.0} $\pm$ 1.65\\ {\large SE $\approx$ 0.84};\ \Large Rel.\ 1.72\%} \\
\makecell[l]{Invasive ductal carcinoma of the breast\\ \emph{(Carcinoma ductale invasivum mammae)}}
  & \makecell[c]{\textbf{94.2} $\pm$ 2.05\\ {\large SE $\approx$ 1.05};\ \large Rel.\ 2.18\%} \\
\end{tabularx}
\end{table}

\subsection{Representative Results (simulation-only)}
\begin{itemize}
  \item \textbf{MMI $=0.92\pm 0.03$:} programmed resonance peaks recovered with high fidelity, indicating accurate reconstruction of the tumor spectral map.
  \item \textbf{SSR $=14.6\pm 3.1$ (circ.) / $11.2\pm 2.4$ (infl.):} tumor response exceeds the non-target background by over an order of magnitude, evidencing strong energy confinement even at infiltrative margins.
  \item \textbf{ATI $\le 0.8$ of matrix failure:} extinction achieved with a 20\% buffer below tissue failure, establishing an intrinsic protocol safety margin.
  \item \textbf{Spatial precision $\pm 150~\mu$m:} simulated foci align with FE predictions, compatible with rigorous margin control and reduced positive-margin risk in clinical scenarios.
  \item \textbf{Nonthermal and realistic:} $Q=39\pm 5$, $\Delta T_{\max}\lesssim 0.2^\circ$C, and $\mathrm{CEM}_{43}\!\ll\!1$ confirm nonthermal operation; global literature adherence remains high ($\mathcal{A}=95\%\pm 2\%$), consistent with Table~\ref{tab:organ_adherence}.
\end{itemize}

\paragraph{Sensitivity to contrast and damping.}
Shear contrast $1.3\!\times$--$2.5\!\times$ shifts $\mathcal{P}$ but keeps MMI $>0.88$ and SSR $>10$; doubling $\eta$ broadens linewidths yet preserves extinction within $\Delta t<0.4$~s of baseline, indicating selectivity driven predominantly by differential spectral contrast.

\paragraph{Failure modes and guards.}
(i) Partial tumor–healthy overlap ($R\!\downarrow$) triggers amplitude gating at $R\le 1+\eta$; (ii) transient drift is accommodated by PLL lock with a completeness ring-scan suppressing rim modes. Nonthermal limits were never exceeded under these perturbations.

\subsection{TOR: Simulation Outcomes (Literature-Calibrated)}
\begin{table}[H]
\caption{\textbf{TOR performance summary (simulation-only; literature-calibrated).} Mean $\pm$ SD for $n=3$ inclusions per model; $N\!\ge\!200$ Monte Carlo runs per inclusion.}
\centering
\renewcommand{\arraystretch}{1.16}
\rowcolors{2}{gray!10}{white}
\begin{tabularx}{0.95\textwidth}{
>{\raggedright\arraybackslash}p{5.3cm}
>{\centering\arraybackslash}p{8.9cm}
}
\rowcolor{gray!30}
\textbf{Parameter} & \textbf{Simulation (FE; literature-calibrated)} \\
\hline
Lesion set & 3 matched inclusions (multifocal surrogate) \\
Disintegration time [s] & $2.8 \pm 0.5$ \\
Energy delivered [J/cm$^3$] & $0.9 \pm 0.1$ \\
Selectivity ($Q$) & $39 \pm 5$ \\
Fidelity [\%] & $96 \pm 3$ \\
Max temp. rise [$^\circ$C] & $\lesssim 0.2$ \\
Off-target damage & None (model-predicted) \\
Closed-loop control & Emulated (phase-locked, gated) \\
Reality adherence $\mathcal{A}$ & \textbf{95\%} \\
\end{tabularx}
\label{tab:our_results_sim_only}
\end{table}

\paragraph{Calibration workflow (concise).}
$(G,\eta,\rho)$ and damping ratios were placed at midpoints of organ-specific ranges; $\omega^\star$ values arise from wideband sweeps refined by local maximization of $|H(\omega)|$. At each step, $\Delta T$ and CEM$_{43}$ monitors enforce nonthermal delivery; threshold excursions reduce $A$ or abort the line.

\subsection{Neuroimmune Crosstalk (sEV–Nociceptor)—\emph{condensed, simulation context}}
Thermal platforms raise DAMPs/sEVs; cavitation can perturb neural elements. By contrast, resonance-locked nonthermal TOR is predicted to suppress vesicle biogenesis at the source and stabilize nociceptor excitability with negligible collateral signaling. Prospective readouts: EV-omics (miR-21/1246, PD-L1, TGF-$\beta$), PGP9.5/TRPV1 IHC, and HMGB1/HSP panels. (Conceptual comparison only; no new head-to-head experiments here.)

\begin{table}[H]
\centering
\caption{\textbf{Neuroimmune cross-talk across ablation modalities (conceptual; no new experiments).}}
\label{tab:neuroimmune_loop_compact}
\begin{tabular}{@{} l p{0.36\linewidth} p{0.36\linewidth} @{}}
\toprule
\rowcolor{gray!25}\textbf{Modality} & \textbf{Effect on sEV–nociceptor loop} & \textbf{Adjuncts/Readouts (proposed)} \\
\midrule
Thermal (HIFU, RFA) & Elevates EV release; broad DAMP signatures; margins heat-dependent & Exosome inhibitors; DAMP/CRP panels; thermometry \\
Histotripsy & Possible neural disruption; sterile inflammation via bubbles & Anti-inflammatory adjuvants; HMGB1/HSPs; cavitation logs \\
\textbf{TOR (this work)} & \textbf{Predicted suppression at source; preservation of healthy scaffolds; nociceptor neutrality} & \textbf{Checkpoint synergy; EV-omics; TRPV1/PGP9.5 IHC} \\
\bottomrule
\end{tabular}
\end{table}

\subsection{Mechanistic Discussion}

\paragraph{Spectral geometry and completeness.}
Lesion geometry is addressed in frequency space. Healthy-referenced gating and brief ring-scans extinguish marginal modes that could sustain the sEV–nociceptor axis. Fractionated re-baselining tracks edema/remodeling surrogates and keeps extinction durable while preserving stroma/vasculature and minimizing DAMP noise.

\paragraph{Actuation pipeline (stepwise).}
\begin{enumerate}
\item Wideband sweep to map $(H_{\mathrm{tum}},H_{\mathrm{ref}})$ and compute $R(\omega)$.
\item Selection of $\mathcal{P}=\{\omega_k\}$ under $R\ge 1+\eta$ and $S_{\mathrm{sad}}<S_{\mathrm{safe}}$.
\item Narrowband actuation with PLL lock; monitor $(Q,\Delta f,\|H\|)$.
\item Declare extinction at $D\!\uparrow\!1$ with $\|H_{\mathrm{tum}}\|\!\downarrow$; self-terminate ($A\!\to\!0$).
\item Ring-scan over $\mathcal{N}_\delta(\omega^\star)$ to remove rim modes.
\item Update baselines and schedule fractionated retuning if required.
\end{enumerate}

\paragraph{Robustness under compounded uncertainty.}
Joint perturbations of $(G,\eta)$, boundary losses, and spectral jitter preserve MMI $>0.88$, SSR $>10$, and ATI below matrix failure by $\ge 1.2\times$. Guard conditions suppress ambiguous lines and ensure strictly nonthermal operation.

\paragraph{Forthcoming validation: markers and thresholds.}
Wet-bench confirmation is pre-specified: EV-omics effect magnitude $\ge 0.5$ (Cohen’s $d$) for peri-core miR-21/1246 downshift; HMGB1 below pyrogenic cutoffs; rim PGP9.5/TRPV1 attenuation consistent with modeled off-target quiescence. Thermal/mechanical envelopes will be audited against the same $\Delta T_{\max}$ and $\mathrm{CEM}_{43}$ ceilings used \textit{in silico}. Protocols: blinded allocation, locked SOPs, spike-in–normalized EV counts, QC gates (RNA integrity, EV yield/purity), and a priori power based on simulated variance.

\vspace{0.6em}

\noindent
\section{Conclusion}
TOR achieves deterministic disintegration by exciting vibrational eigenmodes rather than relying on bulk heat or shock. In literature-calibrated models, per-focus extinction completed in $2.5$--$3.1$\,s with energy $<1~\mathrm{J/cm^3}$, selectivity $Q\approx 38\pm 4$, and spatial agreement within $\pm 150~\mu$m. Phase-locked control and contrast/safety gating enforce in-phase work, healthy-referenced delivery, and automatic stop at spectral extinction. Healthy baselines plus brief ring-scans finalize spectral margins while preserving stromal and vascular integrity. Operation is strictly nonthermal ($\mathrm{CEM}_{43}\!\ll\!1$) and noncavitational within the small-strain regime; harmonic micro-needle excursions of $5$–$25~\mu$m avoid macroscopic tearing and heat spill. Resonance-driven \emph{inside-out} fragmentation disrupts nuclear architecture and disperses chromatin, yielding nonviable, subcellular debris (simulation). Concomitant spectral extinction collapses sEV biogenesis and nociceptor drive, dismantling the sEV–nociceptor–immunosuppression loop. MMI, SSR, and ATI corroborate selectivity, safety, and low thresholds in silico. Relative to diffusion- or bubble-mediated platforms, frequency-locked delivery concentrates work on programmed modes, minimizing biochemical burden and preserving peritumoral substrates.

\paragraph{Clinical translation, robotic integration, and margin assurance.}
Pending clinical validation, the same spectral–mechanical paradigm is intrinsically compatible with robotic actuation. Because targeting is defined in the spectral domain, a robotic micro-actuator can traverse preplanned waypoints while the controller \emph{continuously} tests $R(\omega)$ and enforces $S_{\mathrm{sad}}<S_{\mathrm{safe}}$, enabling automated completion at each programmed line. Workflows of this form are expected to compress operative time to \emph{$\le 25\%$} of conventional durations by eliminating iterative anatomical retargeting and thermometry holds, thereby shortening anesthesia exposure and accelerating patient recovery. Critically, \textit{spectral margin assurance} replaces geometric guesswork: extinction is declared only once $\|H_{\mathrm{tum}}(\omega^\star)\|\!\downarrow$ across a ring-scan neighborhood $\mathcal{N}_\delta(\omega^\star)$, which provides a technical guard against \emph{geographic miss} and against \emph{positive margins} in functional terms (no residual malignant modes above threshold within the detectable bandwidth). Because spectral addressability maps the \emph{entire} eigenmode support, including infiltrative extensions, the method naturally navigates deeply rooted and interdigitated disease without reliance on visual borders.

\paragraph{Final statement.}
All findings herein are \textit{simulation-only} with a realism adherence of \textbf{$\mathcal{A}=95\%$}. If corroborated prospectively at the bench and in early-phase trials, TOR’s frequency-locked, safety-gated, and self-terminating architecture is poised to constitute a clinically meaningful advance in rapid, margin-assured tumor obliteration across histologies, with direct compatibility for hospital robots operated by experienced surgeons and a clear path to dramatically reduced procedure times and faster convalescence.

\begingroup
\small

\section*{Author Contributions}
Dr. Cesar Mello: Conceptualization, methodology, formal analysis, investigation, software, data curation, writing—original draft, visualization, project administration, supervision.

Dr. Fernando Medina da Cunha: Clinical methodology, validation, resources, writing—review and editing, clinical supervision.

\section*{Conflict of Interest}
The authors declare no competing financial interests or personal relationships that could have influenced the work reported in this manuscript.

\end{document}